\documentclass[twocolumn,aps,showpacs,nofootinbib,prd,amsmath,amssymb,floatfix]{revtex4-2}
\usepackage{graphicx}
\usepackage{dcolumn}
\usepackage{bm}
\usepackage{slashed}
\usepackage{lipsum}
\usepackage{color}

\newcommand{\be}{\begin{equation}}                  
\newcommand{\ee}{\end{equation}}                    
\newcommand{\dthreek}{\frac{d^3k}{(2\pi)^3}}        
\newcommand{\Tr}[1]{\mbox{Tr}\left\{#1\right\}}     
\newcommand{\Ima}{\ \mbox{Im}\ }                    
\newcommand{\slsh}[1]{\slashed{#1}}                 
\newcommand{\Gstar}{{}^*G}                          
\newcommand{\Deltastar}{{}^*\Delta}                 

\newcommand{\abs}[1]{|#1|}

\begin{document}

\preprint{APS/123-QED}

\title{Relaxation time for the alignment between the spin of a finite-mass
quark/antiquark and the thermal vorticity in relativistic heavy-ion collisions}
\author{
Alejandro Ayala$^{1,2}$,
David de la Cruz$^3$,
L. A. Hern\'andez$^{1,2,4}$,
Jordi Salinas$^1$}
  \address{
  $^1$Instituto de Ciencias Nucleares, Universidad Nacional Aut\'onoma de
    M\'exico, Apartado Postal 70-543, M\'exico Distrito Federal 04510, Mexico.\\
  $^2$Centre for Theoretical and Mathematical Physics, and Department of Physics,
    University of Cape Town, Rondebosch 7700, South Africa.\\
  $^3$Departamento de F\'isica, Escuela Superior de F\'isica y Matem\'aticas
    del Instituto Polit\'ecnico Nacional, Unidad Adolfo L\'opez Mateos, Edificio 9,
    07738 Ciudad de M\'exico, M\'exico.\\
  $^4$Facultad de Ciencias de la Educaci\'on, Universidad Aut\'onoma de Tlaxcala,
    Tlaxcala, 90000, Mexico.}

\begin{abstract}
We study the relaxation time required for the alignment between the spin
of a finite-mass quark/antiquark and the thermal vorticity, at finite
temperature and baryon chemical potential, in the context of relativistic
heavy-ion collisions. The relaxation time is computed as the inverse of
the total reaction rate that in turn is  obtained from the imaginary part
of the quark/antiquark self-energy. We model the interaction between spin
and thermal vorticity within the medium by means of a vertex coupling
quarks and thermal gluons that, for a uniform temperature, is proportional
to the global angular velocity and inversely proportional to the
temperature. We use realistic estimates for the angular velocities for
different collision energies and show that the effect of the quark mass
is to reduce the relaxation times as compared to the massless quark case. 
Using these relaxation times we estimate the intrinsic quark and antiquark
polarizations produced by the thermal vorticity. We conclude by pointing
out that, in spite of the former being larger than the latter, it is still
possible to understand recent results from the STAR Beam Energy Scan when
accounting for the fact that only a portion of quarks/antiquarks come
from the denser and thus thermalized region in the collision.
\end{abstract}

\maketitle


\section{Introduction}

Results from heavy-ion collisions experiments have  contributed
significantly to our understanding of the properties of strongly
interacting matter at high temperature and density. In these reactions,
two atomic nuclei collide at relativistic energies producing a
deconfined state of hadronic matter, the so called quark-gluon
plasma (QGP). Although many properties of this state have been
revealed by means of a number of different probes, it is  also
fair to say that others still remain elusive at large. One of these
has to do with the possibility to create a vortical fluid in
peripheral collisions. Were this to be the case, the most promising
way to elucidate its properties is by means of the alignment of
particle spin to the global angular momentum, which in turn could
be detected measuring a non-vanishing global particle polarization.
This possibility has prompted the search for global polarization of
hadrons, most notably of $\Lambda$ and $\overline{\Lambda}$~\cite{Becattini2017,Csernai,Sorin2016,Sorin2017,Xie,Pang,Sun, Han,Xia,Teryaev,Karpenko,Suvarieva,Kolomeitsev,Xie2,Guo,Ma}. 
The STAR Beam Energy Scan (BES) program has measured the $\Lambda$
and $\bar{\Lambda}$ global polarizations as functions of the
collision energy~\cite{STAR2007,STARNature,STAR2018}
showing that as the latter decreases, the $\bar{\Lambda}$
polarization rises more steeply than the $\Lambda$ polarization. 
This intriguing result motivates the search for a deeper
understanding of the conditions for the relaxation between angular
momentum and spin degrees of freedom and of its dependence on the
collision parameters such as energy, impact parameter, temperature,
and baryon chemical potential.

In a previous work~\cite{Ayala2020}, we have studied the relaxation time
for spin and thermal vorticity alignment in a QGP at finite temperature
$T$ and quark chemical potential $\mu_q=\mu_B/3$, where $\mu_B$ is the
baryon chemical potential. For these purposes, we resorted to the
computation of the quark self-energy where the interaction with thermal
gluons is mediated by a phenomenological vertex that couples the thermal
vorticity to spin. To make matters simpler, we performed the
calculation for massless quarks whose momentum was small compared
to $T$ and/or $\mu_q$. In this work we remove such approximations
and compute the relaxation time for massive quarks with arbitrary
momentum. We show that the effect of accounting for the quark mass
produces that the interaction rate is larger which in turn translates
into a smaller relaxation time, as compared to the massless quark
case. Other attempts to compute the relaxation time using different
approaches have been reported in Refs.~\cite{Kapusta2020,Liu:2020bbd, Shi:2020qrx}.

This work is organized as follows. In Sec. \ref{Sec:interactionrate}
we calculate the interaction rate for a massive quark at finite density and
temperature. In Sec.~\ref{Sec:results} we show the results obtained
for the quark and antiquark relaxation times as functions of temperature,
collision energy and quark intrinsic global polarization. Finally, Sec.
\ref{Sec:conclusions} provides a summary along with a discussion on the consequences of this
calculation for hyperon polarization.

\section{Quark interaction rate at finite density and temperature}\label{Sec:interactionrate}

Consider a QCD plasma in thermal equilibrium at temperature $T$ and quark
chemical potential $\mu_q$. The interaction rate $\Gamma$ of a quark with
four-momentum $P=(p_0,\vec{p})$ can be expressed in terms of the quark
self energy $\Sigma$ as
\be
    \Gamma(p_0) = \widetilde{f}(p_0-\mu_q)\Tr{\gamma^0\Ima \Sigma},
\ee
with $\widetilde{f}(p_0-\mu_q)$ being the Fermi-Dirac distribution. The
interaction between the thermal vorticity and the quark spin is modeled
by means of an effective vertex
\be
    \lambda^\mu_a = g \frac{\sigma^{\alpha\beta}}{2}\overline{\omega}_{\alpha\beta}\gamma^\mu t_a,
\ee
where $\sigma^{\alpha\beta}=\frac{i}{2}[\gamma^\alpha,\gamma^\beta]$ 
is the quark spin operator and $t_a$ are the color matrices in the
fundamental representation. This vertex builds on the ideas discussed in Ref.~\cite{LiangWang:2005} and
the subsequent studies of Refs.~\cite{Becattini:2007sr, Aristova:2016,Pang:2016,Pang,GaoWang:2015}
and references therein. In order to present a self-contained discussion, we hereby spell out the
ingredients needed for the computation, highlighting the new elements as
compared to Ref.~\cite{Ayala2020}. Recall that the {\it thermal vorticity}
is defined as~\cite{Becattini:2015ska}
\begin{eqnarray}
   \overline{\omega}_{\mu\nu}=\frac{1}{2}\left(\partial_\nu\beta_\mu - \partial_\mu\beta_\nu\right),
\label{thervor}
\end{eqnarray}
where $\beta_\mu=u_\mu(x)/T(x)$, with $u_\mu(x)$ the local fluid four-velocity
and $T(x)$ the local temperature. Thermal vorticity is produced in peripheral
collisions where the colliding matter develops a global angular velocity
$\vec{\omega}=\omega\hat{z}$, normal to the reaction plane, that for our
purposes is chosen as the direction of the $\hat{z}$ axis. The orbital
angular momentum is due to the inhomogeneity of the matter density profile
in the transverse plane~\cite{Becattini:2007sr}. For a constant angular
velocity and uniform  temperature, the magnitude of the thermal vorticity
is given by $\omega/T$.

\begin{figure}[t]
 \begin{center}
  \includegraphics[scale=0.45]{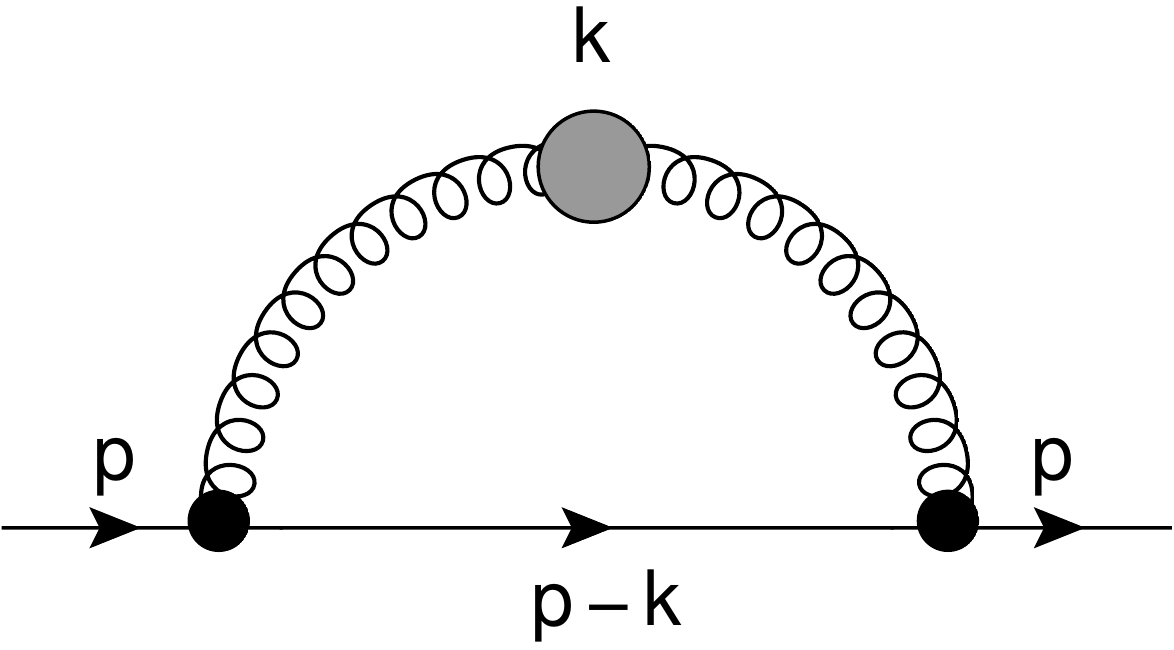}
 \end{center}
 \caption{One-loop
 quark self-energy diagram that defines the kinematics. The gluon line with
 a blob represents the effective gluon propagator at finite density and temperature.
 The blobs on the quark-gluon vertices represent the effective coupling between
 the quark spin and the vorticity.}
 \label{Fig1}
\end{figure}
The one-loop contribution to $\Sigma$, depicted in Fig.~\ref{Fig1}, is
given explicitly by
\be
    \Sigma = T\sum_n \int \dthreek\lambda^\mu_a S(\slsh{P}-\slsh{K})\lambda^\nu_b \Gstar^{ab}_{\mu\nu}(K),
\ee
where $S$ and $\Gstar$ are the quark and effective gluon propagators,
respectively. 
The effective gluon propagator is obtained by summing the geometric series for
the one-loop gluon polarization tensor at high temperature and/or quark
chemical potential. The intermediate quark line is taken as a bare quark
propagator such that the inverse of the interaction rate corresponds to
the relaxation time for the spin and vorticity alignment for quarks that
are originally not thermalized.
\par
In a covariant gauge, the Hard Thermal Loop (HTL) approximation to the
effective gluon propagator is given by
\be
   \Gstar_{\mu\nu}(K)=\Deltastar_L(K)P_{L\, \mu\nu} +\Deltastar_T(K)P_{T\, \mu\nu},
\ee
where $P_{L,T\, \mu\nu}$ are the polarization tensors for three dimensional
longitudinal and transverse gluons, both of which are, of course,
four-dimensionally transverse. The gluon propagator functions for longitudinal
and transverse modes, $\Deltastar_{L,T}(K)$, are given by
\begin{align}
\Deltastar_L(K)^{-1}&=K^2+2m^2\frac{K^2}{k^2}\left[1-\left(\frac{i\omega_n}{k}\right)Q_0\left(\frac{i\omega_n}{k}\right)\right], \nonumber \\
\Deltastar_T(K)^{-1}&=-K^2-m^2\left(\frac{i\omega_n}{k}\right)\Bigg\{\left[ 1-\left(\frac{i\omega_n}{k}\right)^2\right]\nonumber \\
&\quad \times Q_0\left(\frac{i\omega_n}{k}\right)+ \left(\frac{i\omega_n}{k}\right)\Bigg\},
\end{align}
where 
\be
Q_0(x)=\frac{1}{2}\ln{\frac{x+1}{x-1}},
\ee
and $m$ is the gluon thermal mass given by
\be
m^2 = \frac{1}{6}g^2C_A T^2 +\frac{1}{12}g^2C_F \left( T^2+\frac{3}{\pi^2}\mu^2\right),
\ee 
where $C_A=3$ and $C_F=4/3$ are the Casimir factors for the adjoint and
fundamental representations of $SU(3)$, respectively.

The sum over Matsubara frequencies involves products of the propagator
functions for longitudinal and transverse gluons $^*\Delta_{L,T}$ and
the Matsubara propagator for the bare quark $\tilde{\Delta}_F$, such that
the term that depends on the summation index can be expressed as
\begin{eqnarray}
S_{L,T}=T\sum_n\, ^*\Delta_{L,T}(i\omega_n)\tilde{\Delta}_F(i(\omega_m-\omega_n)).
\label{sumprod}
\end{eqnarray}
This sum is more straightforward evaluated introducing the spectral
densities $\rho_{L,T}$ and $\widetilde{\rho}$ for the gluon and fermion
propagators, respectively. The imaginary part of $S_i$ ($i=L,T$) can
thus be written as
\begin{align}
\Ima S_i &= \pi\left( e^{(p_0-\mu_q)/T}+1\right)\int_{-\infty}^{\infty}\frac{dk_0}{2\pi}\int_{-\infty}^{\infty}\frac{dp_0'}{2\pi}f(k_0)\nonumber \\
&\quad \times \widetilde{f}(p_0'-\mu)\delta(p_0-k_0-p_0')\ \rho_i(k_0)\ \widetilde{\rho}(p_0'),
\label{imsum}
\end{align}
where $f(k_0)$ is the Bose-Einstein distribution. The spectral densities
$\rho_{L,T}(k_0,k)$ are obtained from the imaginary part of
$\Deltastar_{L,T}(i\omega_n,k)$ after the analytic continuation
$i\omega_n\rightarrow k_0+i\epsilon$ and contain the discontinuities of the
gluon propagator across the real $k_0$-axis. Their support depends on the
ratio $x=k_0/k$. For $\abs{x}>1$, $\rho_{L,T}$ have support on the (time-like)
quasiparticle poles. For $\abs{x}<1$ their support coincides with the branch
cut of $Q_0(x)$. On the other hand, the spectral density corresponding to
a bare quark is given by
\be
\widetilde{\rho}(p_0')=2\pi\epsilon(p_0')\delta(p_0'^2-E_p^2),
\label{kinrest}
\ee 
where $E_p^2=(p-k)^2+m_q^2$ with $m_q$ the quark mass. The kinematical
restriction that Eq.~(\ref{kinrest}) imposes on Eq.~(\ref{imsum}) limits
the integration over gluon energies to the space-like region, namely,
$\abs{x}<1$. Therefore, the parts of the gluon spectral densities that
contribute to the interaction rate are given by
\begin{widetext}
\begin{eqnarray}
\rho_L(k_0,k)&=&\frac{x}{1-x^2}\frac{2\pi m^2\theta(k^2-k_0^2)}{\left[k^2+2m^2\left(1-\frac{x}{2}\ln \left|\frac{1+x}{1-x}\right|\right)\right]^2+\left[\pi m^2x\right]^2},\nonumber \\
\rho_T(k_0,k)&=&\frac{\pi m^2x(1-x^2)\theta(k^2-k_0^2)}{\left[k^2(1-x^2)+m^2\left(x^2+(x/2)(1-x^2)\ln \left|\frac{1+x}{1-x}\right|\right)\right]^2+\left[(\pi/2)m^2x(1-x^2)\right]^2}.~~~~~~~~~
\end{eqnarray}
\end{widetext}
Collecting all the ingredients, the interaction rate for a massive quark
with energy $p_0$ to align its spin with the thermal vorticity is given by
\begin{align}
    \Gamma(p_0) &= \frac{\alpha_s}{4\pi}\left(\frac{\omega}{T}\right)^2\frac{C_F}{\sqrt{p_0^2-m_q^2}}\int_{0}^{\infty}dk\, k \int_\mathcal{R}dk_0 
    [1+f(k_0)] \nonumber \\
    &\quad \times \tilde{f}(p_0+k_0-\mu_q)\; \sum_{i=L,T}C_i(p_0,k_0,k)\rho_i(k_0,k),
\label{interactionrate}
\end{align}
where $\mathcal{R}$ represents the region
\begin{eqnarray}
k_{0} &\geq& \sqrt{\left(\sqrt{p_{0}^{2}-m_q^{2}}-k\right)^{2}+m_q^{2}}-p_{0}, \nonumber \\
k_{0} &\leq& \sqrt{\left(\sqrt{p_{0}^{2}-m_q^{2}}+k\right)^{2}+m_q^{2}}-p_{0}.
\label{region}
\end{eqnarray}
It can be checked that the region of integration over $k_0$ in Eq.~(\ref{region}) reduces to $-k \leq k_0 \leq k$ when $ m_q\to 0$, as was obtained in Ref.~\cite{Ayala2020}. Notice that  Eq.~(\ref{region}) implies that the available phase space is reduced in the massive quark case, as one
could expect. The polarization coefficients $C_{L,T}$
come from the contraction of the polarization tensors $P_{L,T\, \mu\nu}$ with the trace of
the factors involving Dirac gamma matrices from the self-energy. After implementing
the kinematical restrictions for the allowed values of the angle between
the quark and gluon momenta, these functions are found to be
\begin{align}
    C_T(p_0,k_0,k) &= 8(p_0+k_0)\, \mathcal{C}(p_0,k_0,k), \nonumber \\
    C_L(p_0,k_0,k) &= -8(p_0+k_0)\Big[\mathcal{C}(p_0,k_0,k)-\frac{1}{2}\Big]\nonumber \\
    &\quad -8\frac{p_0\,k^2}{k_0^2-k^2}\,\mathcal{C}(p_0,k_0,k),
\label{polarizationcoeffs}
\end{align}
with
\begin{equation}
\mathcal{C}(p_0,k_0,k) = \left(\frac{k^2-2k_0 p_0-k_0^2}{2k\sqrt{p_{0}^{2}-m_q^{2}}}\right)^2.
\end{equation}
This result should be contrasted with Eq.~(14) of Ref.~\cite{Ayala2020},
which was computed for the massless and small quark momentum limit.
The total interaction rate is obtained by integrating Eq.~(\ref{interactionrate})
over the available phase space
\be
\Gamma = V\int \frac{d^3p}{(2\pi)^3}\Gamma(p_0),
\label{reactionrate}
\ee
where $V$ is the volume of the overlap region in the collision. Recall that,
for the collision of symmetric systems of nuclei with radii $R$ and a given
impact parameter $b$, $V$ is given by
\be
   V=\frac{\pi}{3}(4R+b)(R-b/2)^2.
\label{volume}
\ee
Following the method outlined on Ref.~\cite{XDengHuang2020}, the initial angular
velocity $\omega$ ({\it i.e.}, after full nuclei overlap) produced in Au+Au collisions is
hereby computed performing averages over
\begin{figure}[hb!]
 \begin{center}
  \includegraphics[width=0.48\textwidth]{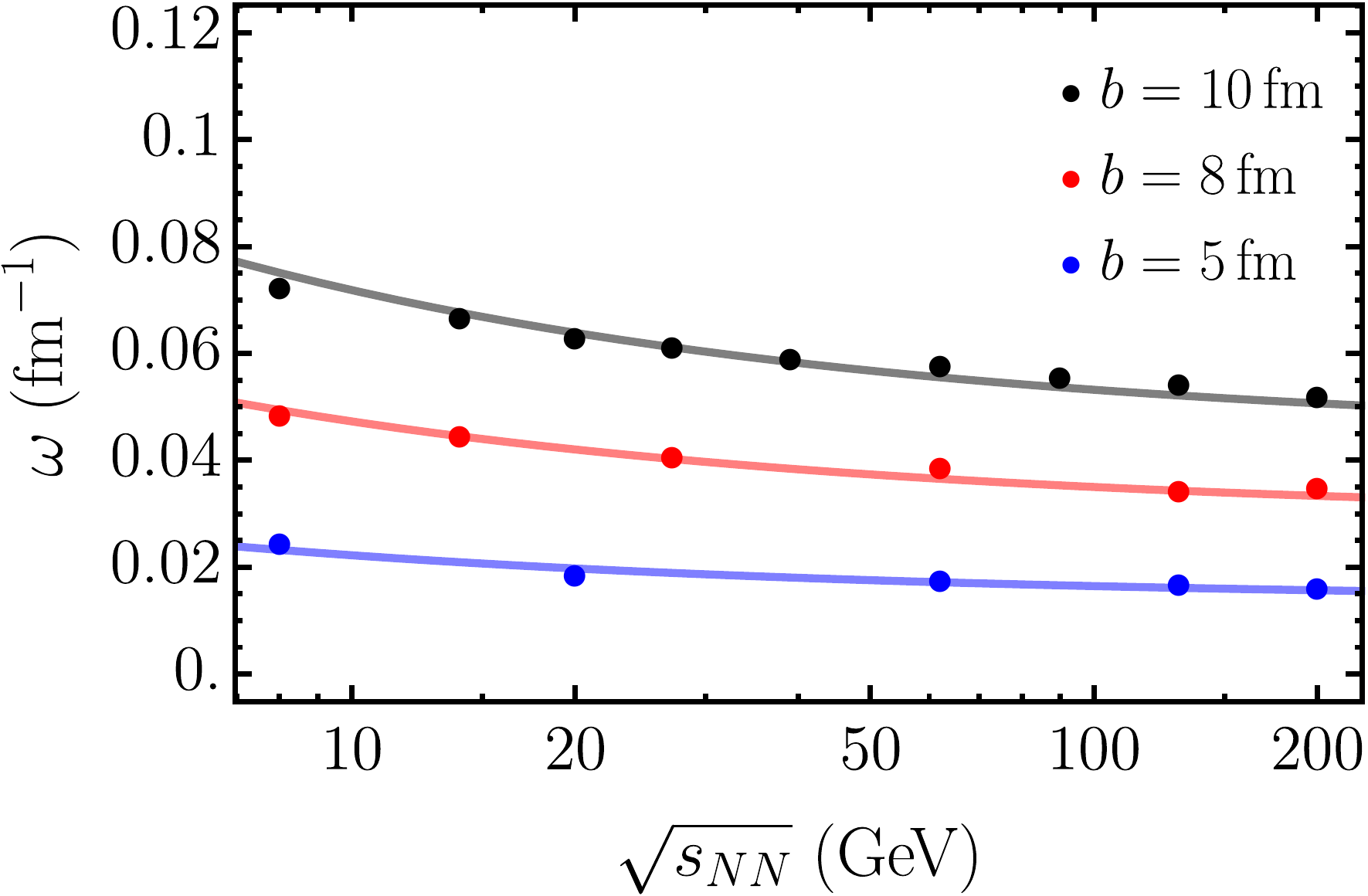}
 \end{center}
 \caption{Initial angular velocity $\omega$ for Au+Au collisions at
 impact parameters $b=5,8,10$ fm as functions of collision energy $\sqrt{s_{NN}}$. Solid lines are the fit of the UrQMD results, using  Eq.~(\ref{fit}).}
 \label{Fig2}
\end{figure}
 $10^4$ events at a given energy and impact parameter using UrQMD~\cite{URQMD}. 
The results of our simulations for the initial angular velocity are shown in Fig.~\ref{Fig2} for three different impact parameters, $b=5,8,10$ fm. Solid lines are the fit of the data using the functional form
\begin{equation}\label{fit}
    \omega = \frac{\omega_0}{2}\, \frac{b^2}{V_N}\left[1+2\left(\frac{m_N}{\sqrt{s_{NN}}}\right)^{1/2}\right],
\end{equation}
where $V_N=(4\pi/3) R^3$, $R=1.1A^{1/3}$, and $\omega_0$ is the 
free parameter of the fit; in all cases $\omega_0 \simeq 1$.
Notice that, as explored in Ref.~\cite{XDengHuang2020}, different choices of smearing functions for the calculation of
the velocity profile, or the definition of the  velocity of the produced particles itself, can lead to a variation on the 
values resulting for the angular velocity; for a more detailed study, see Refs.~\cite{XGHuang:2016,Jiang:2017}.
From the expression for $\Gamma$ in Eq.~(\ref{reactionrate}),
we study the parametric dependence of the relaxation time for spin and vorticity
alignment, defined as
\be
\tau \equiv 1/\Gamma.
\label{relax}
\ee
We now proceed to present the results for the quark and
antiquark relaxation times as computed with Eq.~(\ref{relax}) as functions
of temperature and collision energy as well as the intrinsic quark global
polarization as a function of time.
\section{Results and discussion}\label{Sec:results}
\begin{figure}[b]
 \begin{center}
  \includegraphics[width=0.46\textwidth]{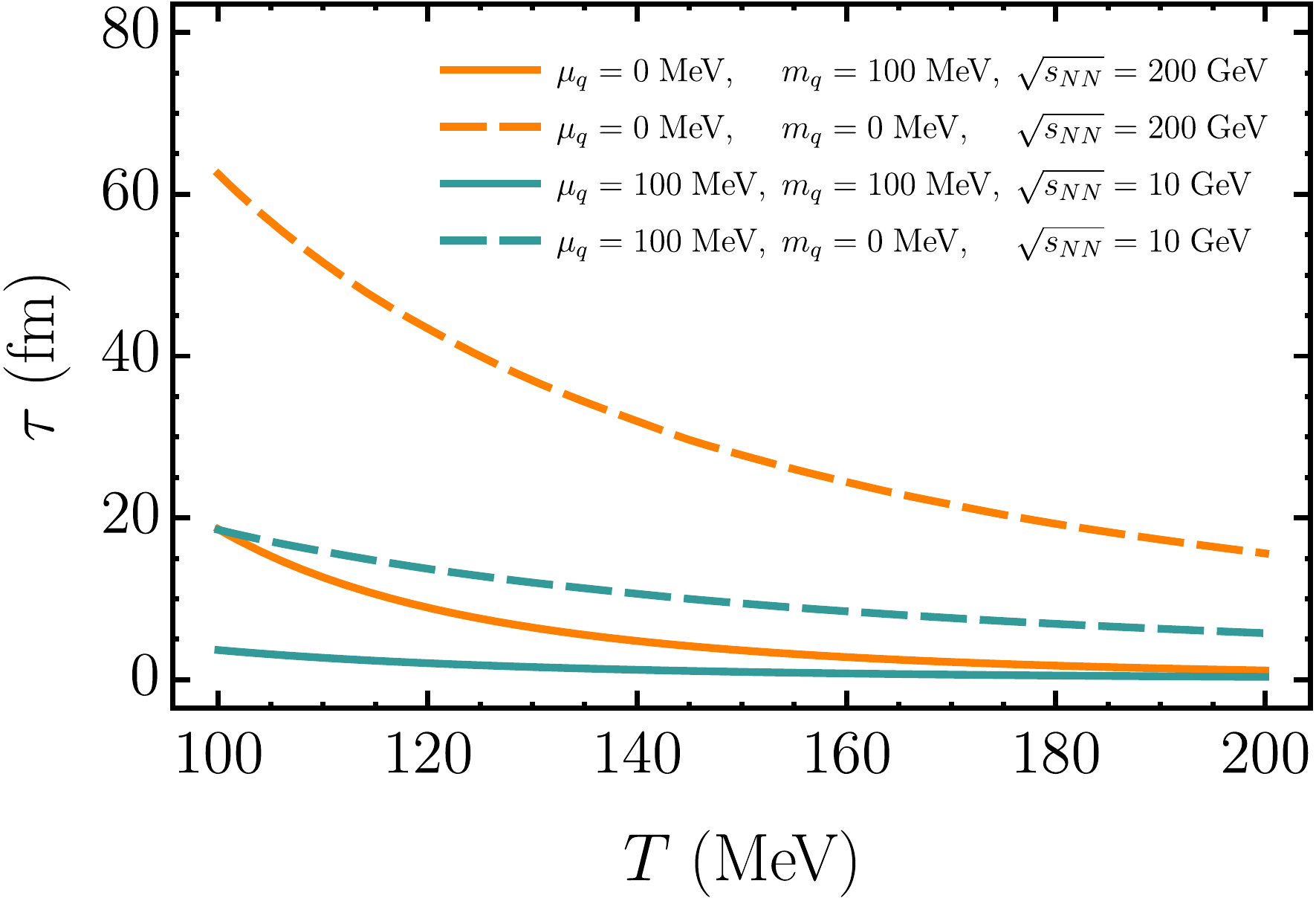}
 \end{center}
 \caption{Relaxation time $\tau$ for quarks as a function of temperature $T$ for semicentral collisions at an impact parameter $b=10$ fm for
 $\sqrt{s_{NN}}=10, 200$ GeV with $\omega \simeq 0.072, 0.051$ fm$^{-1}$, respectively. In dashed lines massless quarks~\cite{Ayala2020}, in solid lines massive quarks.}
 \label{Fig3}
\end{figure}
\begin{figure}[h]
 \begin{center}
  \includegraphics[width=0.46\textwidth]{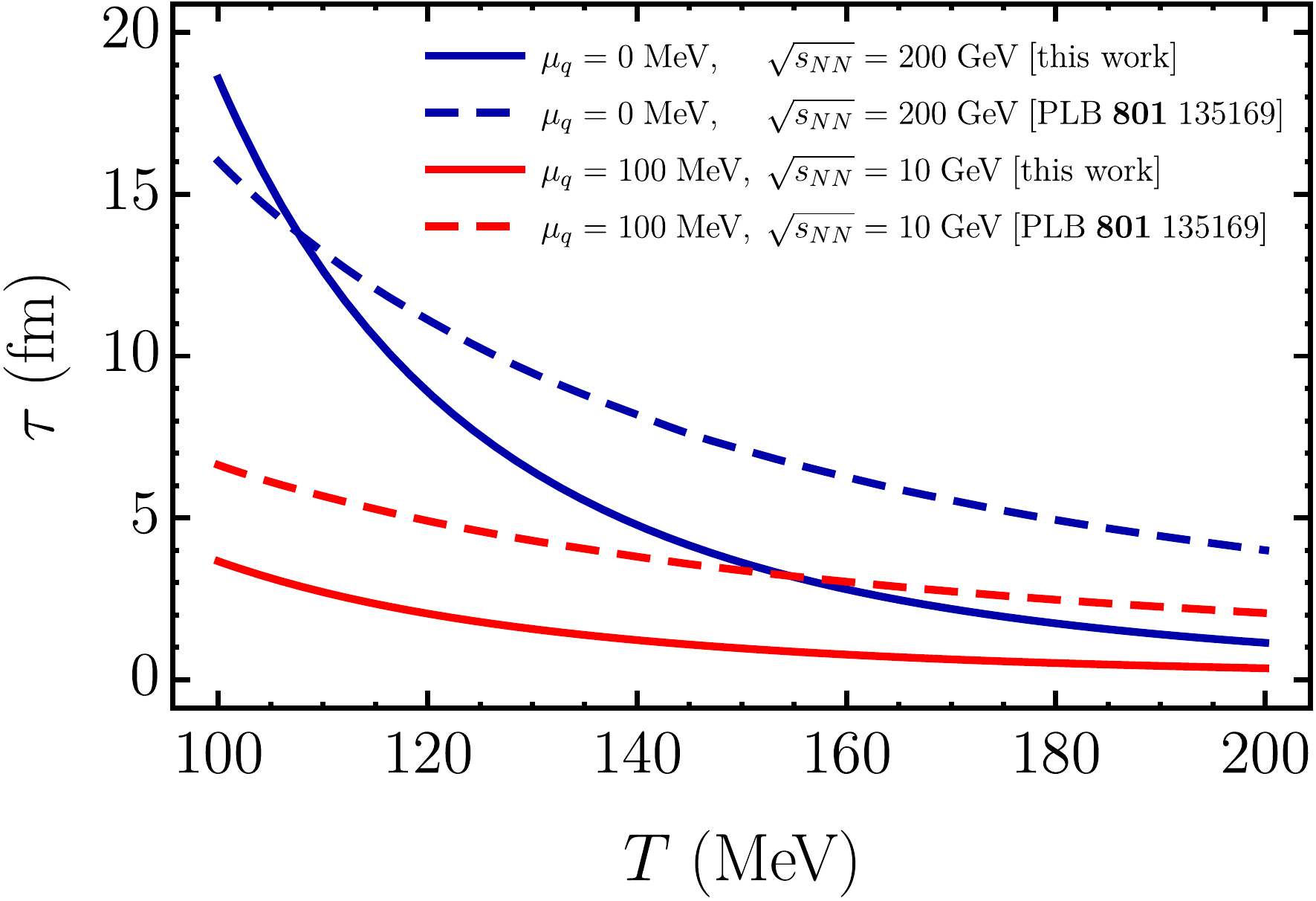}
 \end{center}
 \caption{Relaxation time $\tau$ for quarks as a function of temperature
 $T$ for semicentral collisions at an impact parameter $b=10$ fm for $\sqrt{s_{NN}}=10,200$ GeV
 with $\omega\simeq0.12, 0.10$ fm$^{-1}$, respectively. In
 dashed lines massless quarks~\cite{Ayala2020}, in solid lines massive quarks.}
 \label{Fig4}
\end{figure}
Figure~\ref{Fig3} shows the relaxation time $\tau$ for massive quarks with $m_q=100$ MeV,
corresponding to the strange quark mass, contrasted with the massless quarks case, as a function of temperature $T$, for semicentral collisions
at an impact parameter $b=10$ fm for two different values of quark
chemical potential $\mu_q$ and collision energy $\sqrt{s_{NN}}$. In both cases, $\omega$ is
computed using our simulations described above. The effect of quark mass and of relaxing the assumption of a small quark energy is to reduce the relaxation time for the entire range of considered temperatures, as compared to
the massless case. This may seem counter-intuitive given that
a finite quark mass reduces the available phase space. However, notice that the
new terms in Eq.~(\ref{polarizationcoeffs}), as compared to Eq.~(14) of
Ref.~\cite{Ayala2020}, compensate this reduction and contribute significantly
to a higher interaction rate. Figure~\ref{Fig4} shows the temperature dependence of the relaxation time for
quarks, for two different collision energies and quark chemical
\begin{figure}[b]
 \begin{center}
  \includegraphics[width=0.46\textwidth]{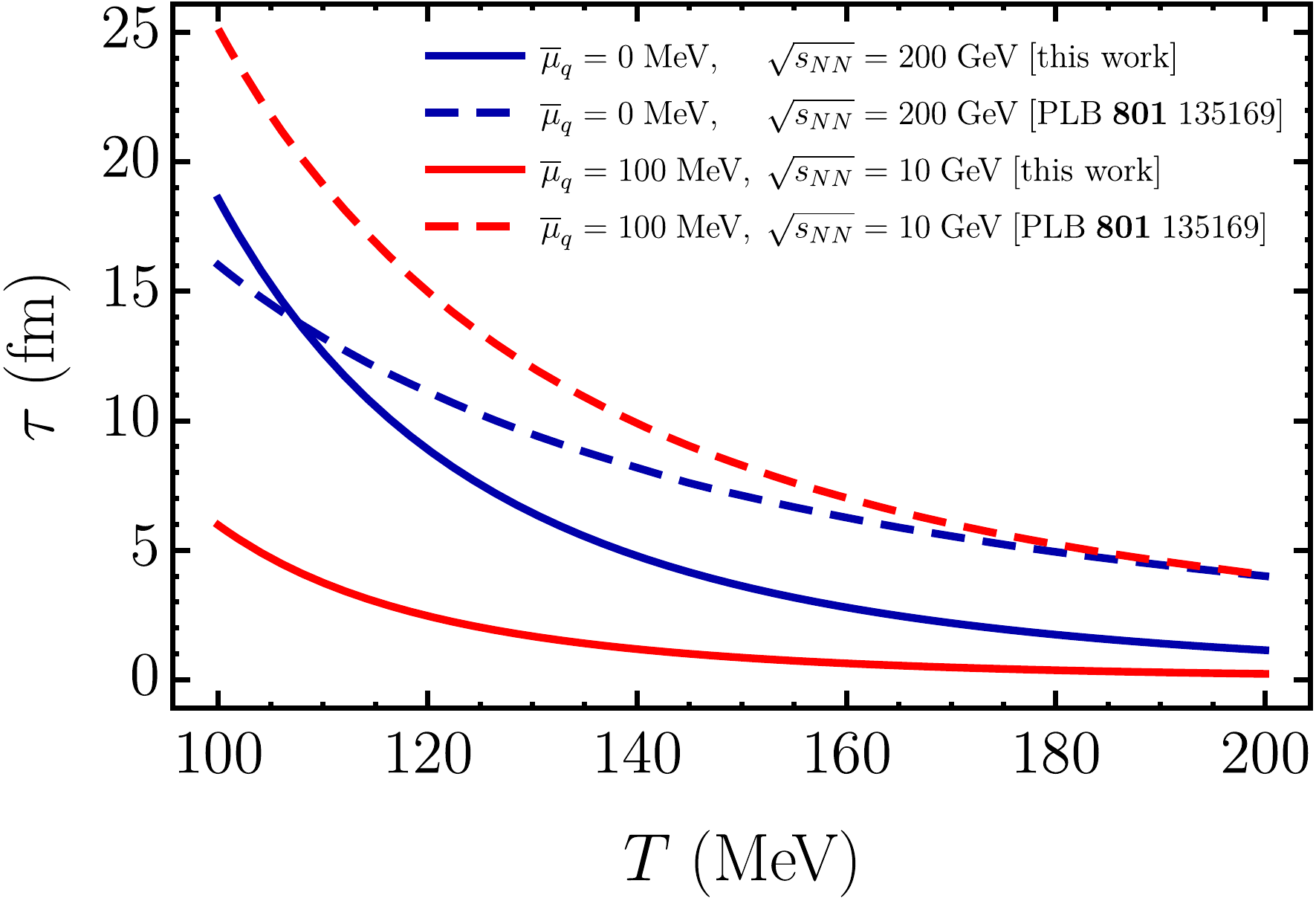}
 \end{center}
 \caption{Relaxation time $\bar{\tau}$ for antiquarks as a function of
 temperature $T$ for semicentral collisions at an impact parameter $b=10$ fm for $\sqrt{s_{NN}}=10,200$ GeV
 with $\omega\simeq0.12, 0.10$ fm$^{-1}$, respectively.
 In dashed lines massless quarks~\cite{Ayala2020}, in solid lines
 massive quarks.}
 \label{Fig5}
\end{figure}
potentials at an impact parameter $b=10$ fm. For direct comparison to
Ref.~\cite{Ayala2020}, the values of $\omega\simeq0.12, 0.10$ fm$^{-1}$ are
used for the massless case. Notice that $\tau\lesssim 5$ fm
for the temperature range 150 MeV $<T<200$ MeV, where the phase transition is
expected to occur. In this temperature range, the relaxation times
are smaller than the ones found in Ref.~\cite{Ayala2020}.\par
Figure~\ref{Fig5} shows the temperature dependence of the relaxation time for
antiquarks (with antiquark chemical potential $\overline{\mu}_q = -\mu_q$) for two different collision energies and chemical potentials at
an impact parameter $b=10$ fm. Again, we use for the antiquark mass $m_q=100$ MeV,
corresponding to the mass of the strange antiquark. The relaxation times for
antiquarks for the temperature range 150 MeV $<T<200$ MeV satisfy $\bar{\tau}\lesssim 5$ fm
and are also smaller than the corresponding relaxation times found in Ref.~\cite{Ayala2020}.
\begin{figure}[th!]
 \begin{center}
  \includegraphics[width=0.46\textwidth]{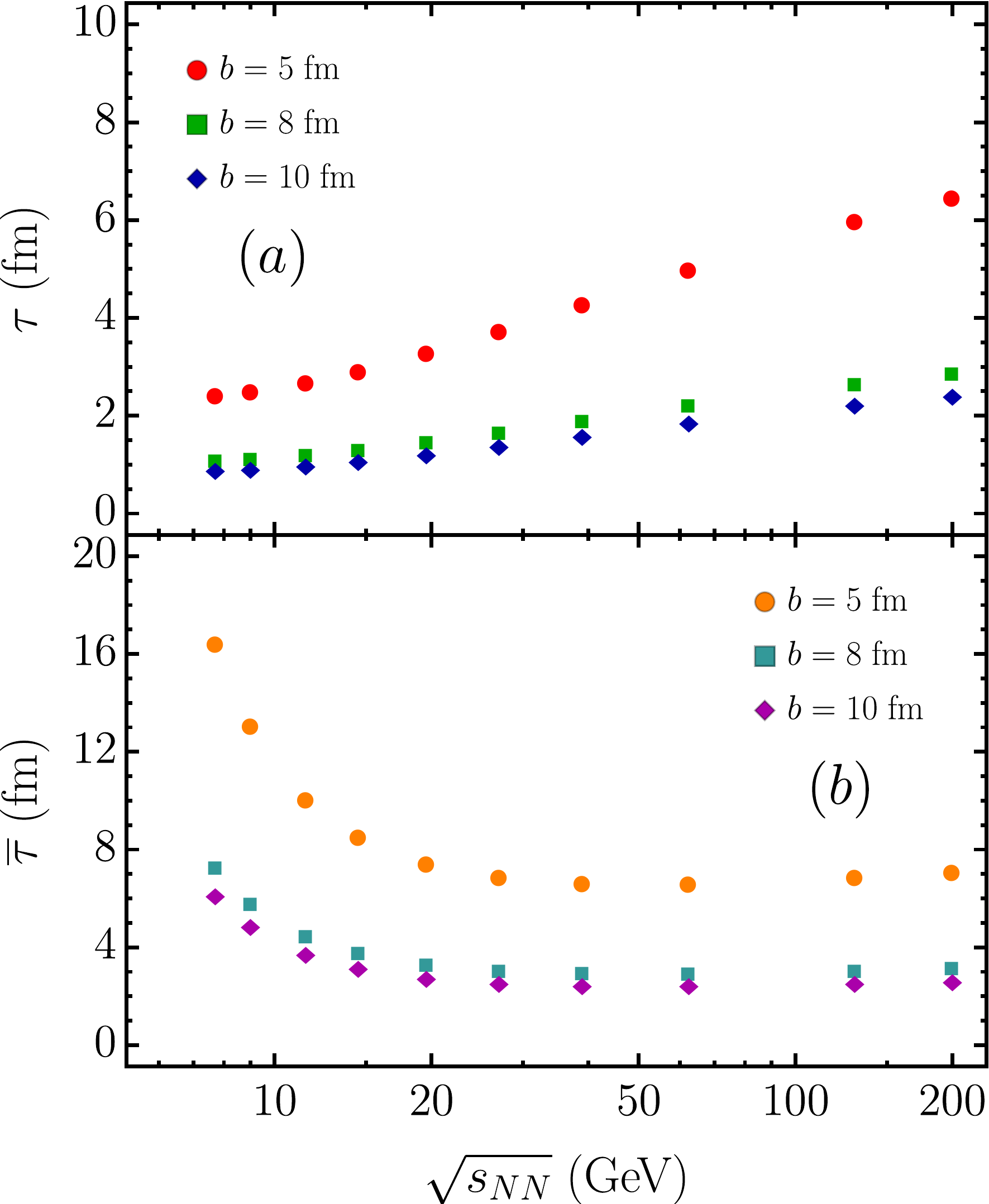}
 \end{center}
 \caption{(a) Relaxation time $\tau$ for quarks as a function of $\sqrt{s_{NN}}$
 for semicentral collisions at impact parameters $b=5,8,10$ fm. (b) Relaxation
 time $\bar{\tau}$ for antiquarks as a function of $\sqrt{s_{NN}}$ for semicentral
 collisions at impact parameters $b=5,8,10$ fm.}
 \label{Fig6}
\end{figure}
\begin{figure}[bh!]
 \begin{center}
  \includegraphics[width=0.46\textwidth]{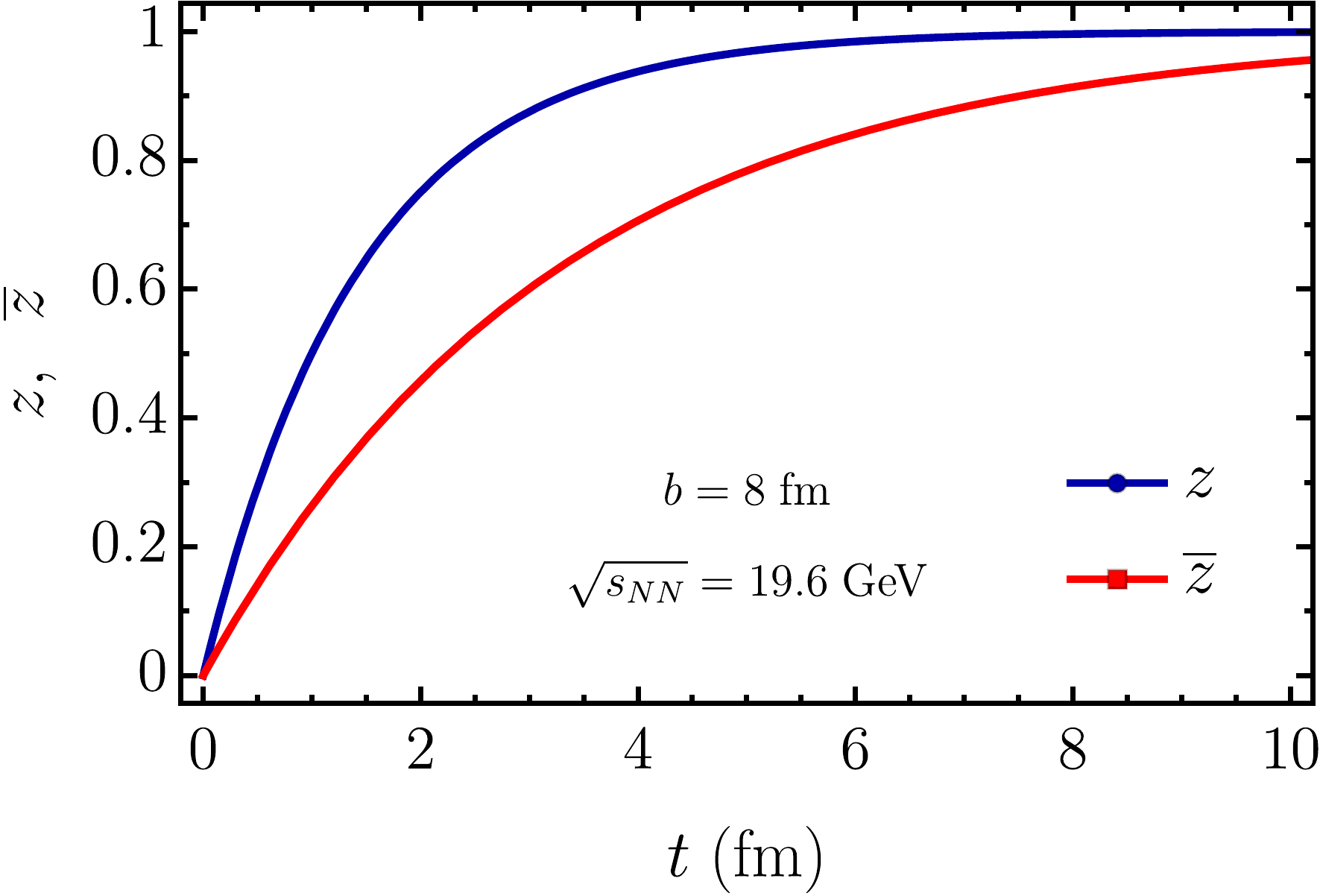}
 \end{center}
 \caption{Intrinsic global polarization for quarks ($z$) and antiquarks ($\bar{z}$)
 as functions of time $t$ for semicentral collisions at an impact parameter $b=8$ fm
 for $\sqrt{s_{NN}}=19.6$ GeV. Notice that $\bar{z}<z$, however, both intrinsic
 polarizations tend to 1 for $t\simeq 10$ fm.}
 \label{Fig7}
\end{figure}
Figure~\ref{Fig6} shows the relaxation time ($a$) $\tau$ for quarks and ($b$)
$\bar{\tau}$ for antiquarks as functions of $\sqrt{s_{NN}}$ for semicentral
collisions at impact parameters $b=5,8,10$ fm as a function of
the collision energy. For each value of $\sqrt{s_{NN}}$, the temperature
$T$ and maximum baryon chemical potential $\mu_B=3\mu_q$ at freeze-out 
were extracted from the parametrization of
Ref.~\cite{CleymansPRC2006} such that
\begin{eqnarray}
T(\mu_B)&=&166 - 139\mu_B^2 - 53\mu_B^4,\nonumber\\
\mu_B(\sqrt{s_{NN}})&=&\frac{1308}{1000+0.273\sqrt{s_{NN}}},
\label{cleynm}
\end{eqnarray}
where $\mu_B$ and $T$ are given in MeV.
Notice that the relaxation times for quarks show a monotonic growth as a function of the
collision energy throughout the energy range considered. In contrast, the corresponding relaxation times for antiquarks
have a minimum for collision energies in the range 40 GeV $\lesssim\sqrt{s_{NN}}\lesssim$ 70 GeV and grows slowly thereafter. 

Figure~\ref{Fig7} shows the {\it intrinsic} global polarization
for quarks ($z$) and
antiquarks ($\bar{z}$), given by
\begin{eqnarray}
   z&=&1-e^{-t/\tau}\nonumber\\
   \bar{z}&=&1-e^{-t/\overline{\tau}},
\label{intrinsic}
\end{eqnarray}
as functions of time $t$ for semicentral collisions at an impact parameter $b=8$ fm,
for $\sqrt{s_{NN}}=19.6$ GeV. Notice that $\bar{z}<z$, however, both intrinsic
polarizations tend to 1 for $t\simeq 10$ fm. From this figure, we also notice that,
even if the QGP phase lasts for less than 10 fm, a finite intrinsic global
polarization, both for quarks and antiquarks, can still be expected.

\section{Summary and Conclusions}\label{Sec:conclusions}

In conclusion, we have used a thermal field theoretical framework
to compute the relaxation times for massive quark/antiquarks (with a mass corresponding to the $s$-quark) whose spin interacts with the thermal vorticity produced in peripheral heavy-ion collisions. With the relaxation times at hand, we also computed the intrinsic quark/antiquark global polarizations.
When this last is
preserved during the hadronization process, one might expect that these
polarizations  directly translate into the corresponding $\Lambda$ and
$\overline{\Lambda}$ polarizations. This would in turn imply that the former
should be expected to be larger than the latter, as opposed to the findings of
Ref.~\cite{STARNature}. Contrary to these expectation, 
Ref.~\cite{AyalaCoreCorona} has recently found that these intrinsic polarization can give rise to the $\Lambda$ and $\overline{\Lambda}$ polarization experimentally observed as a function of collision energy. This result comes from the interplay of $\Lambda$ and $\overline{\Lambda}$ abundances  when their
source in the reaction zone is modelled
as composed of a
high-density core and a less dense corona. Although both regions partake of
the vortical motion, $\Lambda$s and $\overline{\Lambda}$s coming from one
or the other could show different polarization properties as their origins
are different: in the core they come mainly from QGP induced processes,
whereas in the corona they come from n + n  processes. When this fact is
combined with a larger abundance of $\Lambda$s as compared to $\overline{\Lambda}$s
in the corona region together with a smaller number of $\Lambda$s coming from
the core as compared to those coming from the corona --which happens for
semi-central to peripheral collisions-- an amplification effect for the
$\overline{\Lambda}$ polarization can occur. This is more prominent for small
collision energies. More detailed studies of the effect are currently being carried out and will be reported elsewhere.


\begin{acknowledgments}
The authors acknowledge useful conversations with S. Hern\'andez and X.-G. Huang during
the genesis of this work. Support has been received by UNAM-DGAPA-PAPIIT grant number IG100219 and by Consejo Nacional de Ciencia y Tecnolog\'ia grant numbers A1-S-7655 and A1‐S‐16215. 
L. A. H. acknowledges support from a PAPIIT-DGAPA-UNAM fellowship.
\end{acknowledgments}

\end{document}